# Velocity Spectrum Imaging using velocity encoding preparation pulses

By

Luis Hernandez-Garcia, Ph.D.

Alberto L. Vazquez, Ph.D.

Douglas C. Noll, Ph.D.

Corresponding author: Luis Hernandez-Garcia, Ph.D.

hernan@umich.edu

word count: 5097 (Excluding References)

# Abstract

Purpose:

The goal of this article is to introduce a technique to measure the velocity distribution of water inside each voxel of an MR image. The method is based on the use of motion sensitizing gradients with changing first moment to encode velocity. As such, it is completely non-invasive and requires no contrast injections.

Methods:

The technique consists of acquiring a series of images preceded by preparatory RF pulses that encode velocity information, analogously to k-space encoding. The velocity distribution can be decoded via the Fourier transform. We demonstrate its use on a simple flow phantom with known flow characteristics. We demonstrate the technique on the brains of five human participants from whom we collected the velocity distribution along each of the three laboratory axes.

Results:

Velocity distribution measurements on simple phantoms yielded velocity distributions consistent with theory. Human velocity spectra identified specific anatomical features at different velocity bins. The largest fraction of spins was in the lowest velocity bands. Movement in the CSF spaces could be clearly identified at different velocity bands.

Conclusion:

Velocity Spectrum Imaging has great potential as a tool to study the movement of fluids in the human body without contrast agents. In addition to a useful tool for validating computational fluid dynamic models in vivo, it can be used to study the complex movement of water in the glymphatic system and its involvement in neurodegenerative disorders. However, further development is needed to probe the velocity spectrum in the ultra-low velocity regime of the perivascular spaces.

# Introduction

The movement of water in the human body is a very complex system governed by diffusion and convection. These distinct principles can coexist within the same voxel while driving the movement of water molecules within different microscopic compartments at the sub-voxel level. For example, a common occurrence in the brain is that two white matter fiber tracts cross inside the same voxel [1–4]. In that case, there are distinct populations of spins with different diffusion coefficients and directions. Similarly, convective flow occurs not just in the blood, but also along cerebrospinal fluid (CSF) spaces, such as the ventricles and the perivascular space. This convective flow can happen along different directions and velocities within the same voxel, whether it is from capillary blood flow or CSF movement. [5,6]. While diffusion weighted imaging techniques can separate diffusion populations, convective flow imaging techniques (i.e., phase contrast imaging) usually measure an average velocity over the voxel without discriminating different velocity populations inside a voxel [7–10].

In this article, we present, test and implement a new technique to measure the velocity distribution of water inside each voxel of an MR image. This approach is completely non-invasive and requires no contrast agents. The strategy uses modified velocity-selective RF pulses to encode velocity information, analogously to k-space encoding in image formation. The velocity distribution can then be decoded via the Fourier transform. This approach yields the three-dimensional velocity-vector distribution of convective flow in each voxel without differentiating between tissue, blood or CSF signals. Although the original concept of velocity spectrum MR imaging can be traced back several decades [11], the technique has not been fully explored or adopted by the imaging community because of its long acquisition requirements and lack of a clear clinical application.

Here we discuss the theoretical foundation of this method, and demonstrate its use on a simple flow phantom with known flow characteristics. We also demonstrate the technique on human participants and discuss its challenges and potential applications.

# Theory

The theoretical basis for the proposed technique is based on modifications to velocity selective arterial spin labeling (see [12] for a VS-ASL review). Let us consider the velocity selective pulses [13], currently adapted to arterial spin labeling [14,15]. As depicted in Figure 1A, the magnetization is tipped into the *xy*-plane by a $90_y$ pulse. It is then 'flipped' over to the other side of the xy-plane three times by $180_x$ pulses, and finally tipped back to the *z*-axis by a $-90_y$ pulse. A set of gradient blips is introduced between the RF pulses such that the total gradient $0^{th}$ moment is zero, while the first moment is non-zero. This means that immediately before the last pulse segment (Figure 1A, orange arrow), the magnetization vector's phase in the *xy*-plane should be zero if the spins are stationary, but not if the spins are moving at a constant velocity. In that case, their phase is proportional to their velocity and to the first gradient moment (depicted in figure 1B) according to classical Bloch equation analysis.

Specifically, the phase (on the *xy*-plane) of an isochromat traveling at a given velocity (*v*) is predicted by the applied gradient waveform, *G(t)*, and the spins velocity and initial position, $x_0$, , as

$$\varphi = \int \gamma \cdot G(t) \cdot (x_0 + v \cdot t)dt \qquad [1]$$

If we assume constant velocity during the pulse, this can also be expressed as $\varphi = m_g v$ , where $m_g$ is the gradient's **first** moment.

As depicted in Figure 1C, when we tip the isochromat back up with the last -90y segment, the resulting Mz magnetization will be proportional to the projection of the $M_{xy}$ vector onto the *x*-axis (as is done in the Velocity selective saturation ASL[14,16,17]). If we were to use a $-90_x$ pulse to tip the magnetization back up, the resulting $M_z$ component would be the projection onto the *y*-axis.

The original velocity selective pulses used the first-order gradient moment to cause spins traveling over the range of velocities within the arteries (assumed laminar flow distribution) to completely "fan out". When we add the magnetization vectors of the spins inside an artery with a laminar velocity distribution, we find that the net magnetization in the artery after the pulse follows a sinc function of the average velocity in the vessel. As a result, the magnetization in the arteries is nearly canceled when the average speed in the artery exceeds a certain "cut-off" velocity, which is determined by the first gradient moment. When we tip back up with that last segment, the moving spins have near zero net magnetization. This is useful for ASL, as it creates a bolus of saturated blood that can serve as a tracer .

However, in the velocity range below the cut-off velocity, we can also exploit this effect to encode the velocity distribution. After we tip the magnetization back up, the magnetization vector is proportional to the cosine of the accumulated phase in the xy-plane before the tip-up for a single isochromat. More specifically, after the pulse,

$$M_z(v) = M_{xy}(v)\cos(\varphi) = M_{xy}(v)\cos(m_g v) \quad [2]$$

Where $M_z(v)$ is and $M_{xy}(v)$ are the longitudinal and transverse components of the magnetization vector as a function of velocity (v), respectively. The accumulated phase during the velocity selective pulse train is $\varphi$, and $m_g$ is the first gradient moment of the pulse train.

Integrating the magnetization of the spins over a range of velocities for a given gradient moment , $m_g$, yields the net longitudinal magnetization for a given gradient first moment:

$$M_z(m_g) = \int M_{xy}(v)\cos(m_g v)\, dv \quad [3]$$

*We can encode velocity into the phase of the spins in the xy plane ($M_{xy}$ ) by repeating this process multiple times while changing the gradient moment*. Similarly, we can also encode velocity into the sine of the accumulated phase if we tip the magnetization back up using a -90x pulse, instead of a -90y. Combining these two signals as real and imaginary components, the sine and cosine encoded magnetizations result in

$$M_z(m_g) = \int \{M_{xy}(v)\cos(m_g v) + iM_{xy}(v)\sin(m_g v)\}dv \quad [4]$$

This equation is the same as a Fourier Transform of $M_{xy}(v)$, the transverse magnetization as a function of velocity, i.e., the spectrum of spin velocities in the voxel.

The implication is that, by sampling the signal multiple times with multiple gradient moments, we obtain a Fourier encoded distribution of the velocity. We can then decode the transverse magnetization as a function of velocity, $M_{xy}(v)$, by using the inverse Fourier Transform.

It is important to note that this spectrum is proportional to the distribution of convective flow velocities (or velocity density, $\rho(v)$) at a microscopic level *within each voxel*, even though the spatial resolution of the images will be determined by the readout portion of the pulse sequence.

The velocity range and resolution that can be encoded is determined analogously to k-space encoding of spatial information. Assuming uniform sampling of the velocity spectrum by stepping through a range of gradient amplitudes (and thus first gradient moment values, $m_1$ ). The maximum velocity ($V_{max}$) that can be encoded is determined by the step size between velocity encoding gradient moments, $\Delta m_1$, as

$$V_{max} = \frac{\pi}{\gamma \Delta m_1} \quad [5]$$

and the velocity resolution, $\Delta V$, is determined by the largest velocity encoding first gradient moment, $m_{1,max}$, as

$$\Delta V = \frac{\pi}{\gamma m_{1,max}} \quad [6]$$

Analogously to the relationships between field of view and spatial resolution and k-space sampling in image formation.

To our knowledge, there is little work investigating the use of velocity spectrum imaging. The original concept of velocity density encoding was proposed by Moran in 1982 [11] and Wong et al recently leveraged this concept to measure arterial pulsatility by using velocity encoding bipolar gradients during the readout [18]. Our lab introduced an alternative strategy using velocity selective pulses in 2015 in the context of displacement encoding[19]. The present article expands on this later strategy to resolve velocity distributions at the sub-voxel level. The practical challenges posed by this measurement include primarily cardiac pulsatility and other bulk motion effects. Furthermore, aliasing from high velocity spins beyond the "cut off velocity" (i.e., arterial flow) can occur.

# Methods

All scanning protocols were carried out on a 3.0 T GE UHP scanner (Waukesha, WI, USA) with a 32 channel receive coil (Nova Medical, Wilmington, MA).

We acquired measurements to determine the feasibility of the proposed method using two custom-built flow phantoms and also conducted preliminary experiments on human volunteers (N=4).

We designed a flow phantom consisting of a spherical water chamber that contained an array of 8 parallel cylinders of diameters from 2.25 mm to 4mm at equal intervals. These cylinders carried water between two separate chambers on opposing sides of the sphere (top to bottom). We positioned the phantom such that the water flowed from top to bottom, to ensure that the pressure difference across all tubes was the same. In this configuration, each cylinder had a different laminar velocity distribution, whose mean velocity was determined by its diameter. A peristaltic pump placed in the scanner control room drove the flow through the phantom using 1.27 cm diameter laboratory tubing. We dampened the pump pulsatility by including a reservoir in line with the pump to absorb the vibrations and trap bubbles in the system. A CAD rendering of the phantom is shown in figure 2. The phantom was 3D printed using an SLA printer (Form 3+, Formlabs, Somerville, MA, USA) and filled with tap water doped with Nickel Chloride.

We first measured the mean velocity in the tubes using the vendor supplied, phase-contrast MRI pulse sequence (TR/TE/FA = 11.1 ms / 4.8 ms/ 8deg, matrix size = 512 x 512, 60 slices, voxel size=0.39x0.39x5.66 mm, acceleration factor = 2), using a velocity encoding gradient equivalent of 30 cm/s along the direction of flow (Anterior-Posterior). Mean tube velocity was obtained by averaging the values in hand-drawn circular regions covering the inner volume of each tube.

We collected a velocity spectrum image series along the flow axis of the tubes (A-P) on the multi-velocity phantom as 30 pairs (sine + cosine projections for each velocity encoding step) of gradient-echo images using a 3D spherical projection spiral readout trajectory (TR/TE/flip=3500, 3.5ms, 15 deg., FOV = 20 cm, matrix = 96 x 96 x96, number of spiral readouts per train = 30, num. interleaves = 4, total imaging time: 14 minutes). The readout was preceded by a cosine velocity encoding pulse (odd image frames) or a sine encoding pulse (even frames) as described in the theory section. The pulses are depicted in figure 1A. We varied the velocity encoding gradient amplitude from -4 G/cm to 3.73 G/cm for each of the velocity-space encoding steps (image pairs). These gradient amplitudes corresponded to encoding 0.58 cm/s resolution from -8.06 to 8.64 cm/s.

We reconstructed the individual images in the series using a Total-Variation regularized, model-based, conjugate-gradient SENSE reconstruction using the Michigan Image Reconstruction Toolbox (MIRT) available at (https://github.com/JeffFessler/mirt). We then obtained velocity spectrum images by (1) rephasing the image series at each voxel, using the zero-encoded frame as reference, (2) combining the cosine and sine encoded images into complex images for each encoding gradient, (3) detrending the image series with a third order polynomial, (4) multiplying by a Hanning window to reduce truncation artefacts, (5) Fourier transforming the image series at each voxel and using the magnitude of the spectrum and (6) Normalizing the spectrum (ie – scaling it such that the integral of the spectrum equals 1) to calculate the spin density fraction at each velocity.

The second flow phantom consisted of a simple loop of laboratory tubing (1.27 cm diameter) submerged in a container of $CuSO_4$ doped water. As a result, the flowing water has opposite velocity directions on opposite sides of the loop. The pump and this “loop phantom” are shown on figure 4. The same pump and damping system as above controlled the flow through this phantom.

We collected a velocity spectrum image series on the loop phantom as before, albeit with the following modifications. The velocity encoding gradients were varied from 0 to 4 G/cm (half velocity k-space sampling) over 31 pairs of images (cosine and sine encoded). Since the velocity spectrum is expected to be real-valued, we did not collect negative gradient amplitudes because of the conjugate symmetry of the Fourier Transform of real-valued data.

We also modified the base BIR-8 velocity encoding pulse to increase the first moment of the velocity encoding by increasing the gaps between pulses. The modified pulses allowed us to sample higher values of the velocity k-space using the same gradient amplitude range as before (-4 to 4 G/cm). The new velocity resolution was 0.085 cm/s, and the velocity spectrum width spanned from -2.6 to 2.6 cm/s. We computed the velocity spectrum image series as before.

For reference, we computed the theoretical velocity density (expressed as the fraction of particles moving at each velocity) inside a tube of flowing water under laminar flow conditions for three maximum velocity values. It can be readily shown that the distribution is described by

$$\rho(v) = 2\pi R v \sqrt{1 - \frac{v}{V_{max}}} \qquad [7]$$

Where $\rho$ is the velocity density, $R$ is the radius of the tube, $v$ is velocity and $V_{max}$ is the maximum velocity (at the center of the tube).

Human participants gave informed consent in compliance with the University of Michigan’s Internal Review Board. We collected velocity spectra along each of the laboratory frame’s axes using the same pulse sequence as in the phantom experiments. (TR/TE/flip= variable, 3.5ms, 15 deg., FOV = 22 cm, matrix = 64 x 64 x 64, number of spiral readouts per train = 30, num. interleaves = 2, total imaging time ~8 minutes per axis). However, we used lower resolution and included a cardiac-gated pre-saturation pulse 3 seconds before each encoding pulse to reset the magnetization between encodes and to mitigate cardiac pulsatility effects. The velocity encoding gradients were varied from 0 to 4 G/cm over 31 pairs of images (cosine and sine encoded, Half-Fourier sampling). By taking advantage of the Fourier transform’s conjugate symmetry property, we could fill in the missing negative

Velocity-k-space data) and recover a velocity spectrum width from -2.6 to 2.6 cm/s at 0.085 cm/s resolution. We smoothed the velocity encoded image series using a Savitzky-Golay filter and computed the velocity spectrum image series as before. We reduced the effect of scanner and physiological drifts and the dominance of the stationary water peak by global signal regression. Specifically, we calculated the spatial mean for each image of the velocity spectrum image series and used it to construct a global mean spectrum regressor. We then fitted and removed this global mean regressor from the image series using linear regression.

# Results

Figure 3 shows the theoretical laminar flow velocity distributions scaled as the percentage of particles in the tube moving at each velocity. The distribution is determined by the maximum velocity at the center of the tube. Note that the tube's diameter will not affect the velocity distribution *fraction*, given a maximum velocity within the tube, although the maximum velocity is a function of the diameter along with the pressure gradient, and viscosity of the fluid.

Figure 4 summarizes the results from the parallel flow phantom experiment. A single slice of the phase-contrast derived velocity image is shown superimposed (blue) on a structural image of the phantom in Figure 4A. The panel also shows the *average* velocities in the phantom's tubes calculated from circular ROIs in the phase contrast images. A white square indicates the region used to create the plots in figure 4C. Below, figure 4B shows the velocity spectra (distribution) calculated from ROIs in half of the tubes in the flow phantom (the other half is omitted for clarity). The ROIs were manually selected by identifying the central voxel of each tube in each of the three center slices. The velocity distribution in each tube approximates a skewed parabolic distribution between zero and the maximum velocity in the tube as expected for laminar flow distribution. We note that in the case of higher velocity tubes (7.24 cm/s average), the distribution wraps around (aliases) to the negative side of the velocity spectrum, as predicted by the Nyquist sampling theorem.

Figure 4C shows a different perspective of the same data. Here, we show the 'velocity density' at each of several velocity bins in the spectrum. They have been cropped to show the central region containing the flow tubes in panel A (white square ROI). The velocity bin images have been vertically concatenated and only every other velocity bin is shown for clarity. The zero velocity bin has been scaled down by a factor of 15 for display purposes. As expected, only the fastest tubes can be seen in the faster velocity bin images.

Figure 5 shows the velocity spectrum from an ROI (3x3x3 voxels) inside the tube of the Loop-Phantom, chosen from segments with opposite flow directions and a third voxel from the stationary water chamber. Two different velocity-bin axial images through the tube are displayed on the right panel, reflecting the spectral plots on the left panel: the positive (+0.76 cm/s) and negative (-1.44 cm/s) velocity bins show high velocity densities on opposite sides of the tube, and the zero-velocity voxel shows high intensity throughout the stationary chamber. The method can differentiate between positive and negative flow directions, even when sampling only the positive portion of the velocity k-space. This feature enables us to reduce the scan time by half. However, we also note that the true velocity spectrum outside the tubes should contain only a large peak at zero velocity. However, the detrending step in the computation eliminates the zero velocity peak, as expected by Fourier transform theory, and what remains is largely noise.

The data acquired from human studies successfully captured velocity spectra along the three principal axes at every voxel. The 3D velocity spectrum images from our human participants are available for download at (https:// TBD upon publication) as NIFTI format. The intent of this communication is not a thorough analysis of the velocity distribution patterns in the human brain and thus we will limit the scope of this article to highlighting some observed features of the observed spectra obtained from this preliminary cohort of healthy participants.

.

Figure 6 depicts some observations for one of the participants. The figure depicts orthogonal views of the velocity density fractions at four selected velocity bins along the three velocity encoding axes. They are displayed in a different color for each axis (red=x-axis, green=y-axis, blue = z-axis) with 50% oppacity. The velocity fraction images are thresholded at 2% and overlaid on a T2 weighted image of the participant

In Figure 6A, We observe a significant fraction of water moving upwards the z-axis around the brainstem at the +0.425 cm/s band, and in the frontal part of the ventricles.

Upward movement is also evident in the CSF of the calcarine fissure at + 0.34 cm/s in the sagittal view of Figure 6B. At the same velocity bin, we also observe movement along the x and y axes next in the CSF surrounding the brain stem in the axial and coronal views.

In the -0.17 cm/s velocity band, movement of water is apparent in the cerebral aqueduct, predominantly in the y-axis in the axial view of Figure 6C. Also, sagittal sinus flow is evident at as flowing along the negative y-axis and z-axis directions consistent with anatomical expectation, but also along the x-axis, which is perhaps counter-intuitive. We attribute this movement to pulsatility of the adjacent arteries.

At the -0.935 cm/s band, shown in Figure 6D, we can see the movement of water in a region encompassing the choroid plexus along the x-axis.

Supplementary Figures S1A through S5A shows montages for all five participants of the spin density fractions at three velocity bins (out of 61) along the three encoding axes. The bins are centered at -1.4, 0 and +1.4 cm/s. The images are shown as orthogonal sections. We note that the fraction of water moving a zero velocity is below 1% in the ventricles in four of the five participants. On the other hand, the fraction of water at the higher velocities is noticeably higher in the ventricles than the surrounding tissue. Spatial gradients can be observed in some of the velocity fractions, which we attribute to eddy currents and other other gradient imperfections.

Supplementary Figures S1B through S5B show the middle slice of the whole velocity spectrum series along all three axes for the each participant. For simplicity, the figure only shows every 5$^{th}$ velocity bin between +/-2.175 cm/s (including 0 cm/s velocity bin).

Figure 7 shows velocity spectra extracted from cubic regions of interest (3x3x3 voxels) selected based on anatomy. The regions were extracted from the sagittal sinus, cerebral aqueduct, right and left motor cortices, frontal white matter regions and both of the frontal ventricular regions. We observe that white and grey matter regions have largely symmetric spectra, but the ventricles and acqueduct show asymmetric and heterogeneous velocity distribution patterns in the low velocity regime, as expected. The spectrum from the sagittal sinus is also largely asymmetric along all three axes, reflecting the direction of flow. As noted above, the presence of a x-direction velocity components is counterintuitive and we attribute it to pulsatility from the adjacent meningeal arteries.

# Discussion

This article introduces proof of principle of a method to image the distribution of velocity within living tissue using velocity selective labeling pulses to encode a velocity spectrum. The concept of imaging the velocity spectrum dates to 1982, [11] but there has been little progress in this area until now. The method extends the principles of phase contrast imaging to encode the entire velocity spectrum into a series of images that can be decoded to recover the fraction of spins moving at each of the velocities in the spectrum. Mathematically, it follows a similar formalism as phase encoding in MRI image collection.

In this report, we have verified that the method can capture velocity distribution spectra by scanning two simple phantoms with known flow characteristics and then tested the method on three human volunteers with consistent results. The proposed method was able to distinguish and measure velocity distributions in three dimensions.

The results from the phantom experiments were consistent with our predictions. The results from the human experiments indicate the complex flow pattern in the brain at the low velocity range. We can consistently identify anatomical features at different velocity bands, such as the cerebral aqueduct and the sagittal sinus, as expected.

The velocity of CSF water in the ventricles has a complex distribution, oscillating with the cardiac cycle predominantly in the range of 0-1.5 cm/s, but with average values less 0.4 cm/s in the ventricles [20] . Prior measurements in animals have found capillary velocities to be pulsatile and in the order of 0.1 – 0.2 cm/s [2122]. Water in the perivascular spaces is also pulsatile and has been measured in rodents to be in the range of 0.001 - 0.004 cm/s [23], fluid mechanical models for humans predict velocities of 0.006-0.0260 cm/s [24]. Thus, our data suggest that the proposed method is suited to examine the movement of water in the slow regime of the CSF spaces, but resolving the velocity distribution in the perivascular space will require significant further development.

This method may prove to be a powerful tool to study the movement of water in biological samples non-invasively and to validate fluid mechanical models. Potential applications include the study of the complex movement of water in the glymphatic system and its involvement in neurodegenerative disorders. Further investigation is needed to understand the full utility of the approach and characterize the normative distribution of the velocity spectrum in the human brain.

Additionally, this technique can be a useful tool for validating computational fluid dynamic models in vivo.

Other methods to image the glymphatic system non-invasively are centered on measuring the glymphatic exchange by the use of spin preparations. A recent method targets the exchange between blood and CSF, particularly at the choroid plexus. Another strategy uses arterial spin labeling and differentiates between capillary and extravascular water by leveraging diffusion [25–27] or T2 relaxation [28–30] differences between these compartments. Another recent strategy combines T2 weighting to isolate the CSF signal with velocity encoding to identify glymphatic movement [31,32]. Our method fits into this latter category, although we do not isolate the CSF compartment. However, our approach produces a more nuanced picture of the velocity distribution in the tissue.

**Limitations**

There are several important limitations to the proposed technique, as currently implemented in the present study and further development is needed to address them. First, the method's theoretical limitations are primarily determined by sampling theory: the number of encoding steps and their amplitude determine the velocity bandwidth that can be sampled without aliasing.

We note that our data did not identify any large arteries, because the velocity content of the arterial blood is significantly higher than the velocity spectrum can capture at our current velocity sampling bandwidth. Since velocity distribution from laminar flow in the arteries spans a wide velocity range above the sampling velocity bandwidth (e.g., see figures 3 and 4), we can expect the arterial velocity distribution to be aliased multiple times and 'spread' out continuously throughout the entire velocity spectrum rather than a specific, identifiable, velocity band.

Fast laminar flow results in cancellation of the magnetization of the flowing spins when the fastest flowing isochromats acquired more than $2\pi$ radians of phase during the velocity encoding pulse. Note that this phenomenon belies the principle of velocity selective saturation ASL [12,14,15] If one is interested in examining the vasculature with the proposed method, a broader velocity spectrum needs to be sampled. Other phase-contrast [33] or ASL methods [34] are better suited for this purpose.

Eddy currents and field imperfections in BIR-8 pulses were studied in depth in the development of the original BIR-8 pulses by Guo and Meakin [12,14]. Among several designs of velocity selective pulses, the symmetric BIR8 design was found to be the most robust to B0 and B1 inhomogeneity, and least sensitive to eddy current effects. Mitigation was achieved by inclusion of gaps between gradient and RF pulses. Although we also included such gaps in the design of our own pulses, these effects are still be apparent in our data, particularly along the z-axis (see supplementary figures S1 through S5). We can still observe spatial gradients at some velocity bands that are consistent with eddy currents, and not anatomy. Further work is necessary to reduce these effects further by pulse design and/or image post-processing.

A limitation of the proposed technique is that the velocity encoding process also encodes diffusion information into the observed signal. This is expected because the velocity encoding gradients in the preparatory pulses are quite similar to those used in diffusion imaging prep pulses [3536]. While we have neglected this effect in this preliminary report, the proposed method encodes both diffusion and velocity and are not clearly separable. Fortunately, the diffusion effects are relatively minor, as the b-values of the preparation pulses are relatively small (0 to 25 s/mm$^2$), but this is a limitation of the proposed technique that will necessitate further development.

The pulsatile nature of flow in perivascular spaces and other CSF compartments, such as the cerebral aqueduct can also potentially confound the proposed method. For example, if the periodicity of back-and-forth movement of the aqueduct were very fast (within the ~45 ms. duration of the encoding pulses) it would severely affect the amount of phase accumulated during the encoding pulse and result in erroneous phase encoding. However, in the case of the slower periodicity associated with the human heartbeat, this back-and-forth flow is slower than the encoding and readout periods and can thus be mitigated by cardiac gating, although not completely eliminated. In our implementation, we did not model acceleration or target a specific cardiac phase, which is a limitation that still needs to be addressed in future work.

It should also be noted that our technique does not differentiate between CSF and blood. It only gives a velocity distribution over all compartments. Isolation of a particular compartment could be achieved, but may require additional preparation nulling pulses (e.g., VASO techniques as in [37,38]) . Our assumption, based on previous literature [25-27], is that the CSF water in the perivascular space is significantly slower, while capillary blood water moves faster. However, this assumption limits the interpretability of the data.

An important limitation is the duration of the scans, as presented in this work ( > 7 minutes per axis), which is acceptable only for demonstrating proof of principle but is not practical for clinical use. Our experiments employed a minimum of two interleaves per image and per encoding step and the interval between frame acquisitions was a minimum of 4 seconds in the human data.

Sparse sampling techniques in space and velocity can and will be further explored to achieve faster scanning rates. Similarly, scan timing parameters must also be optimized in subsequent work.

It is unclear whether the proposed velocity spectra yield information about glymphatic flow in the parenchyma and perivascular spaces at this point, given the coarse resolution of our images, and that the lowest velocity we can resolve is 0.095 cm/s. Future work will focus on increasing our velocity resolution and investigating whether the fraction of water at the slowest velocity bin is informative about the glymphatic movement.

# Figures

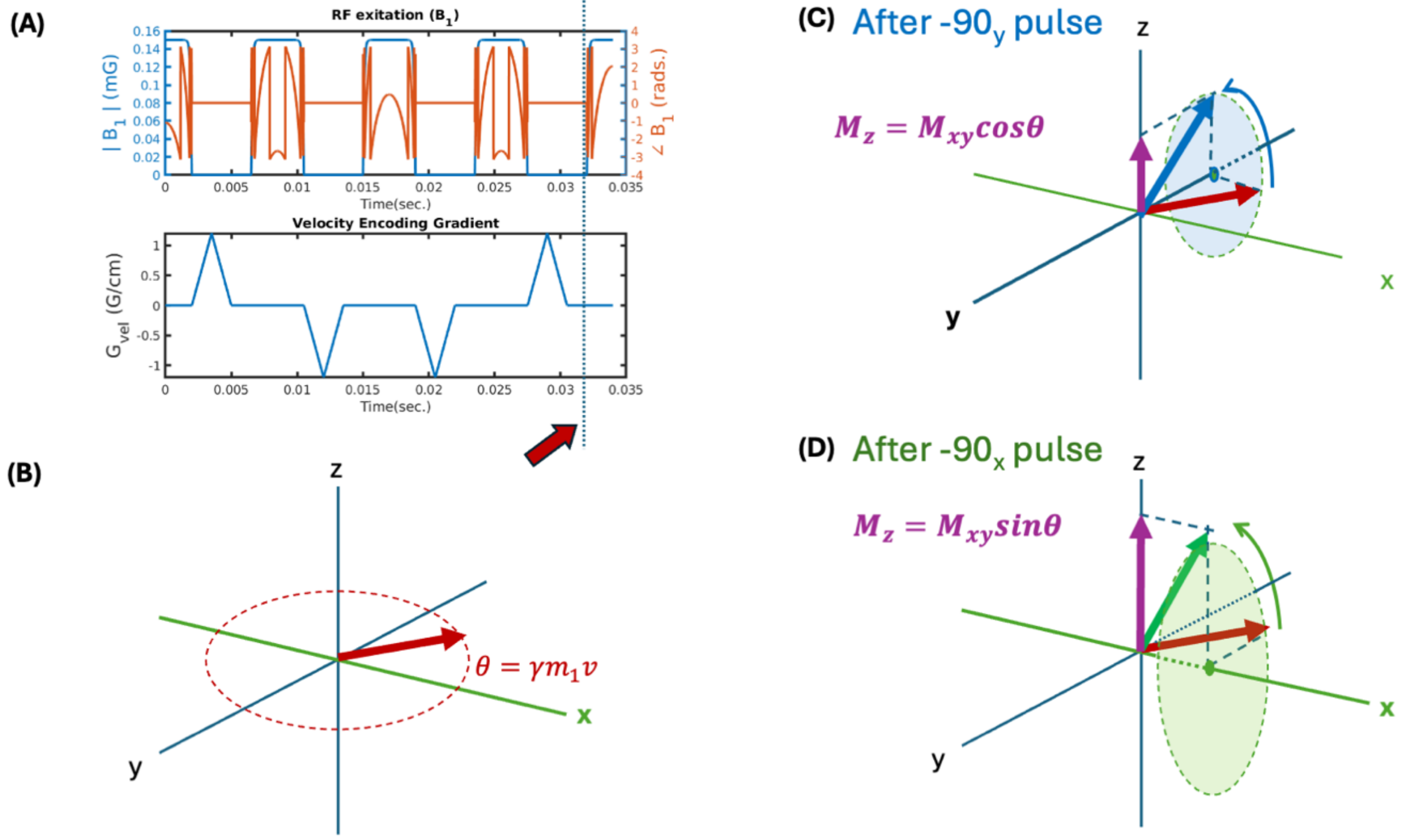

Figure 1. Velocity Encoding Pulse Train and its effect on moving isochromats (A) Diagram of RF and Gradient waveforms. (B) Magnetization of an isochromat traveling at velocity v immediately before the last RF segment of the velocity encoding pulse train (C) Effect of the last RF segment, rotating the magnetization about the Y axis (D) Similarly, modification of the last segment such that the rotation happens about the X-axis.

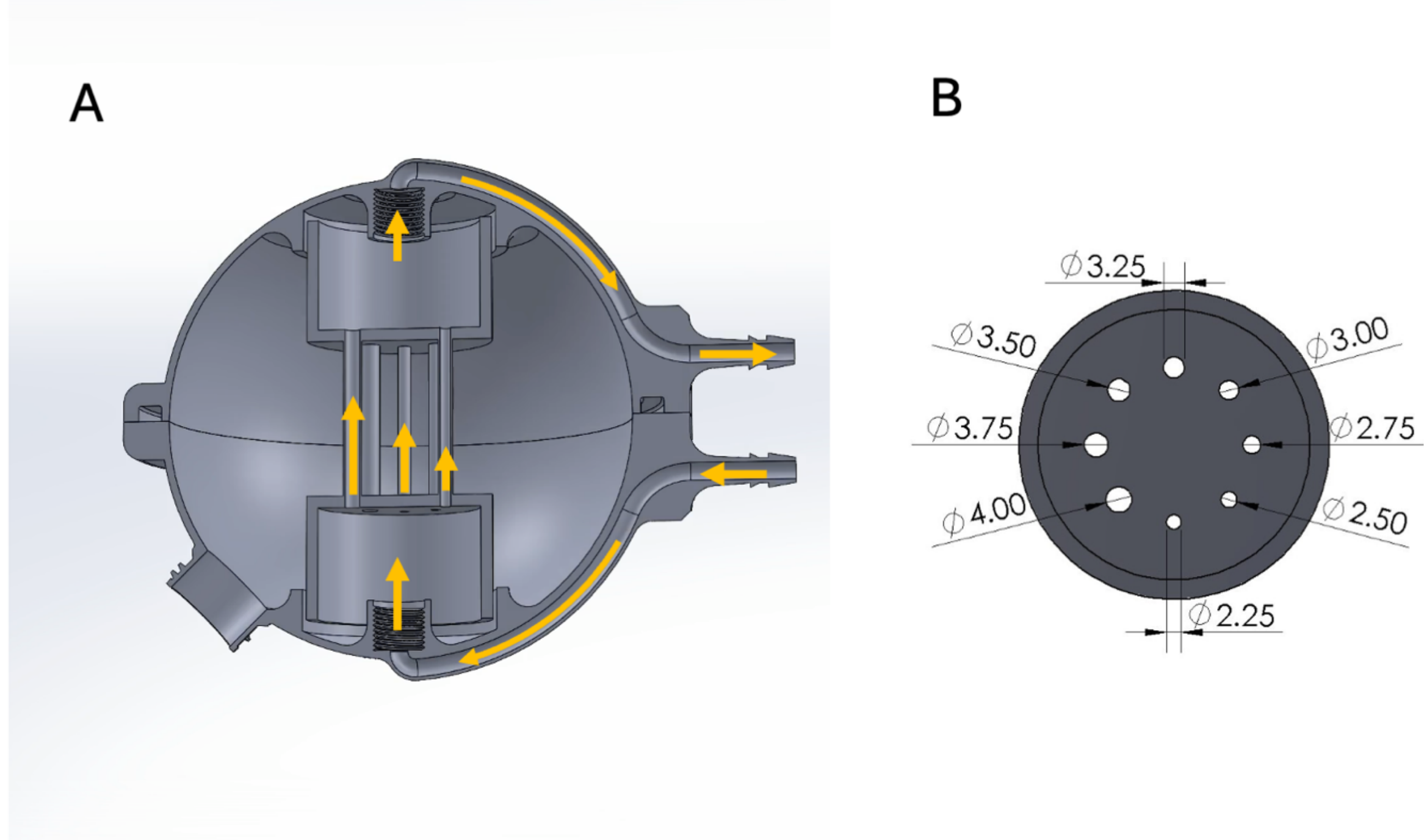

Figure 2. The multi-velocity phantom. (A) cross section view through the phantom. It is comprised of a spherical chamber for stationary water and an internal chamber made of multiple tubes of varying diameters. This internal chamber is fed through a larger chamber and drains into a second chamber. The pressure drop across all the tubes is the same. Fluid flow is provided by a circular peristaltic pump connected to the inlet and outlet on the right side of the panel. (B) A cross section of the internal flow chamber indicating the tube internal diameters in mm.

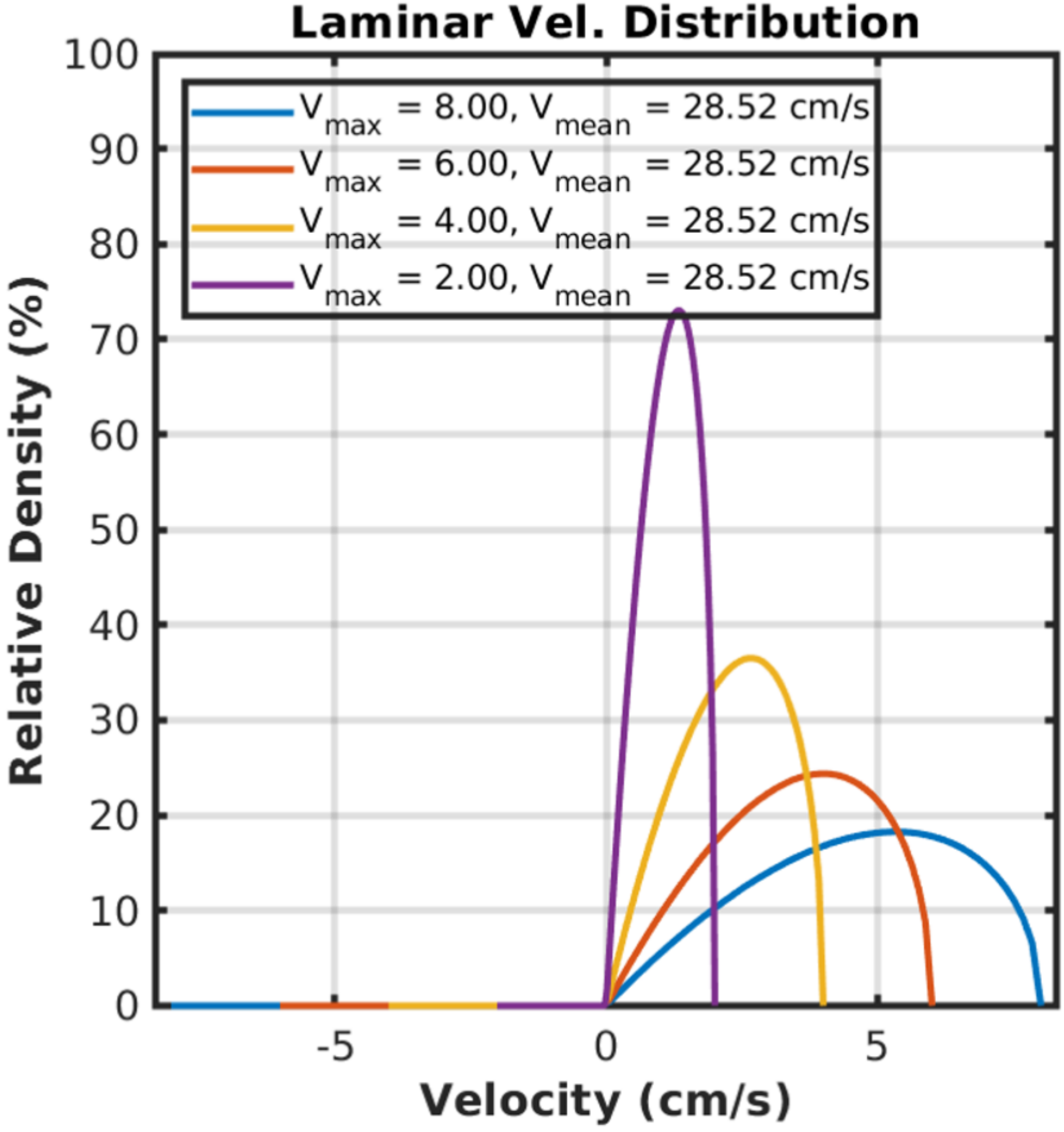

Figure 3. Simulated Laminar Flow Velocity Distributions. The velocity distribution is expressed as the percentage of particles in the tube moving at each velocity. The distribution is a skewed parabola determined by the maximum velocity at the center of the tube. The tube's diameter will not affect the velocity distribution fraction, given a maximum velocity within the tube. However, the maximum velocity is a function of the diameter along with the pressure gradient, and viscosity of the fluid.

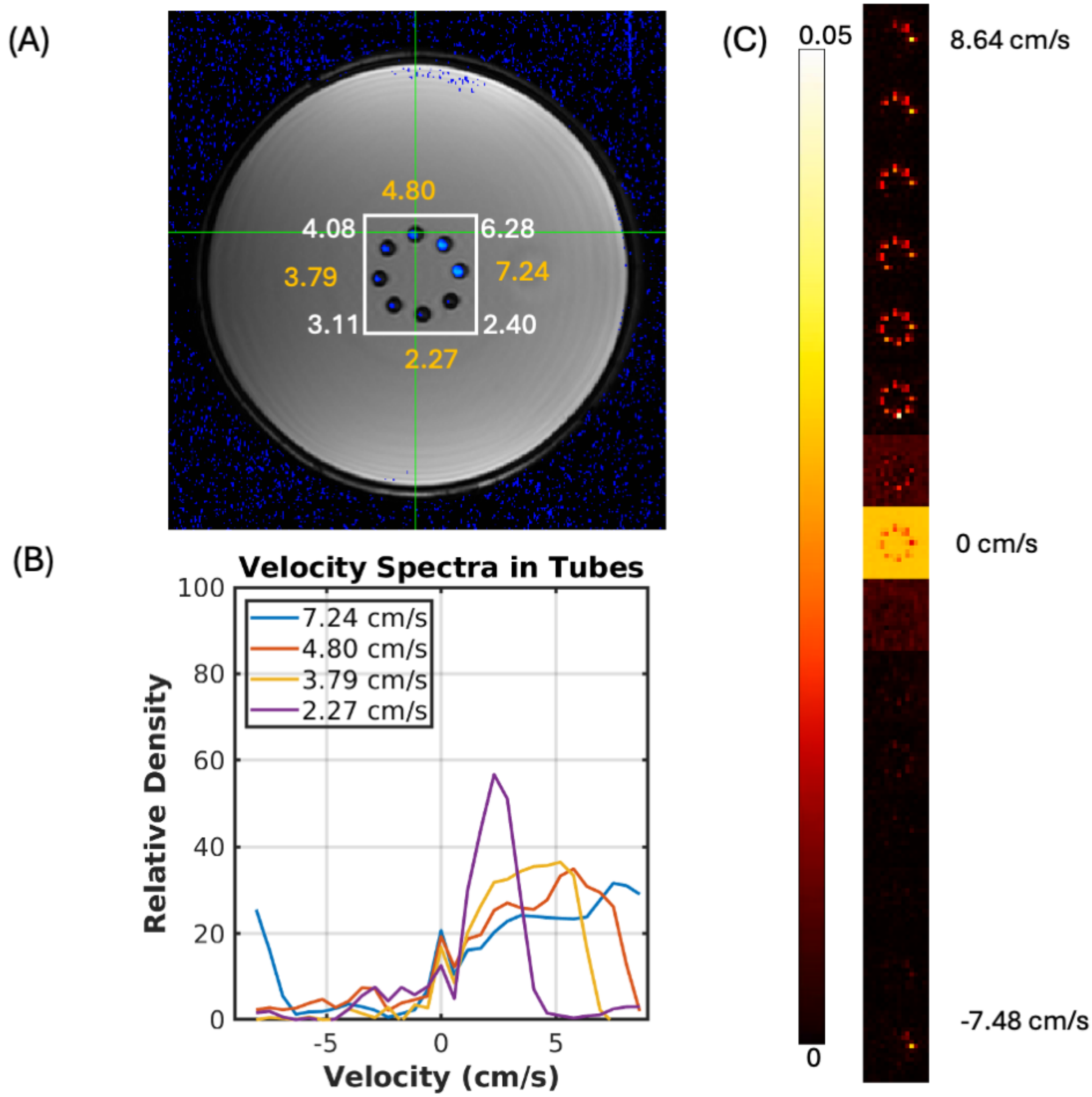

Figure 4. (A) Overlay of the velocity map obtained from phase contrast imaging on top of a structural image of the velocity phantom. The cross section indicated the average velocity in each of the tubes (B) Velocity spectra calculated from every other tube in the phantom expressed as the fraction of spins traveling at each velocity (c). Zoomed images from an ROI encompassing the flow tubes in the phantom. For clarity, we only show the density map at every third velocity bin of the spectrum. The zero-velocity bin has been scaled down by a factor of 15 for display purposes.

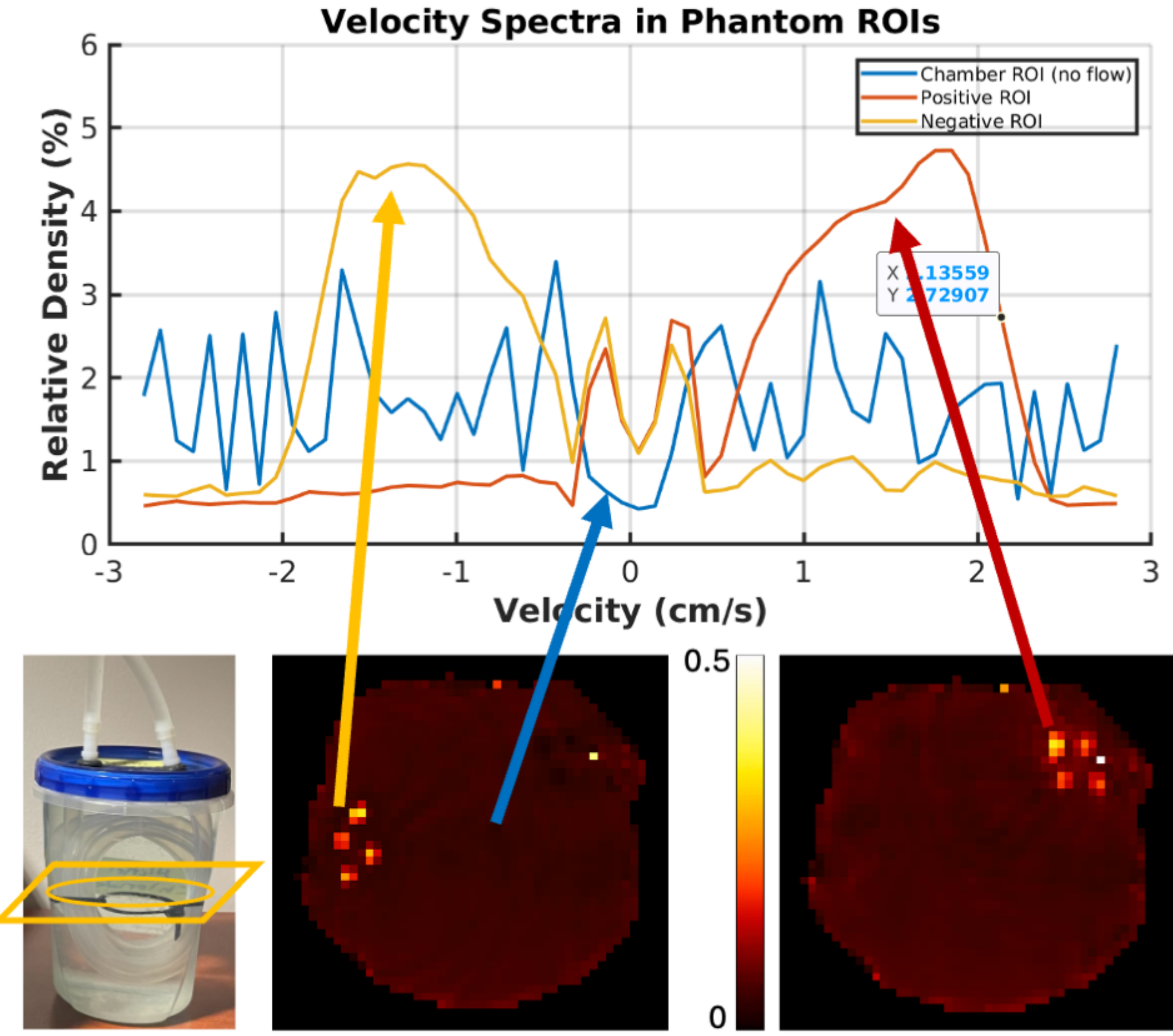

Figure 5. Velocity Spectrum Image Series from the simple Loop phantom verifying the capability of the method to differentiate between flow directions along the same axis. The top panel shows the velocity spectra from three voxels in the image. Note that the spectra are calculated as the fraction of the spins in the voxel traveling at each velocity and plotted here on a log scale. The red line is from a voxel chosen from the tubes on the left side of the phantom, where the water was flowing in the negative direction. The blue line is from a voxel chosen from the main chamber of the phantom, outside of the flow tubes. The yellow line comes from a voxel in the tubes on the right side of the phantom, where the flow is positive. The lower panels depict the phantom on the left and the spin density fraction in a slice through the flowing tubes at -1.44 cm/s (left) and +0.76 cm/s (right)

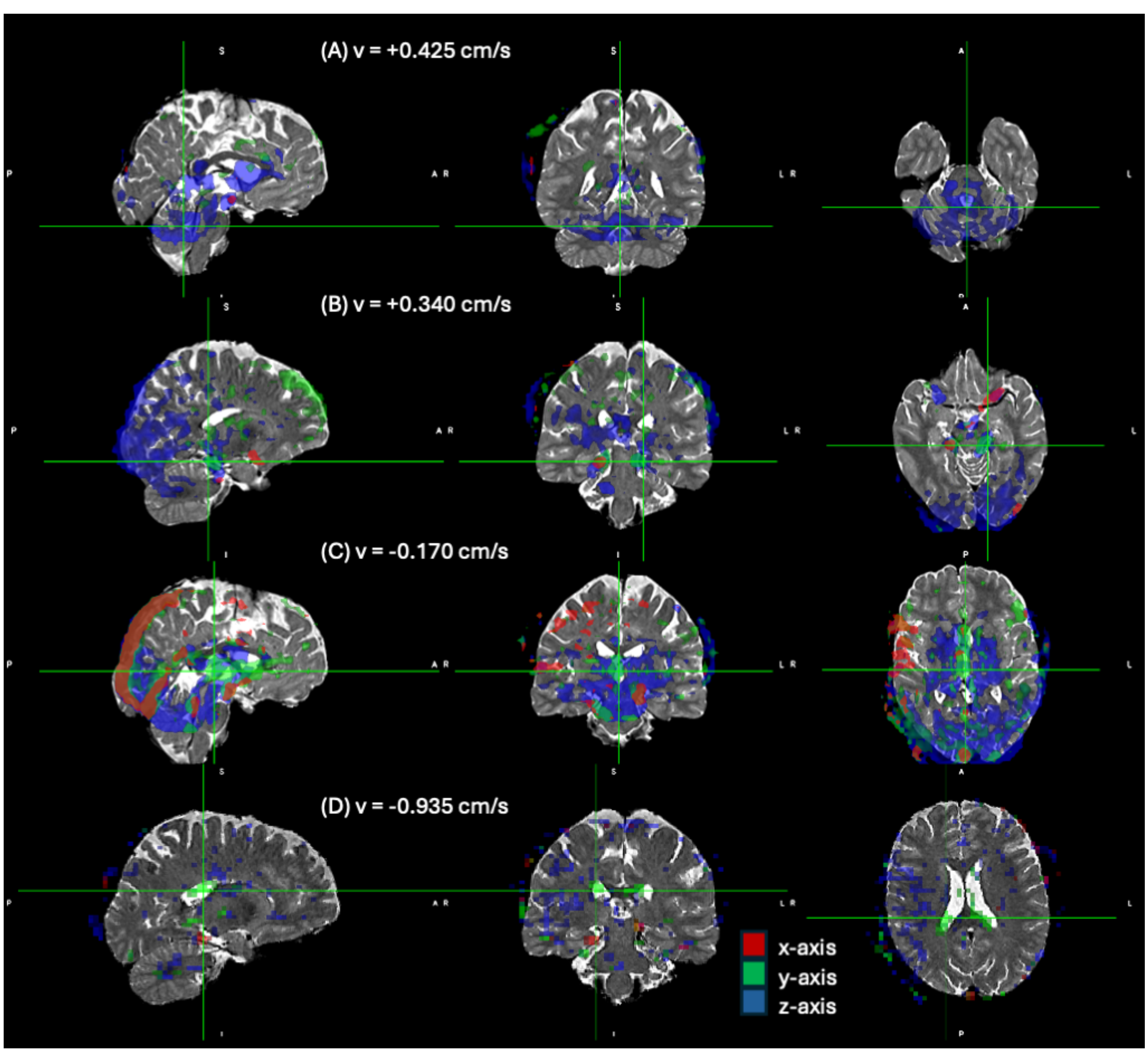

Figure 6. Selected views of the velocity spectrum from one participant. Four orthogonal views of the water fraction at different velocity bands along all three axes are overlaid on a T2 weighted anatomical image. The water fractions are thresholded at 2% in each axis. The three directions are displayed in different colors as indicated, and they are overlaid with 50% opacity to show the overlap between maps.

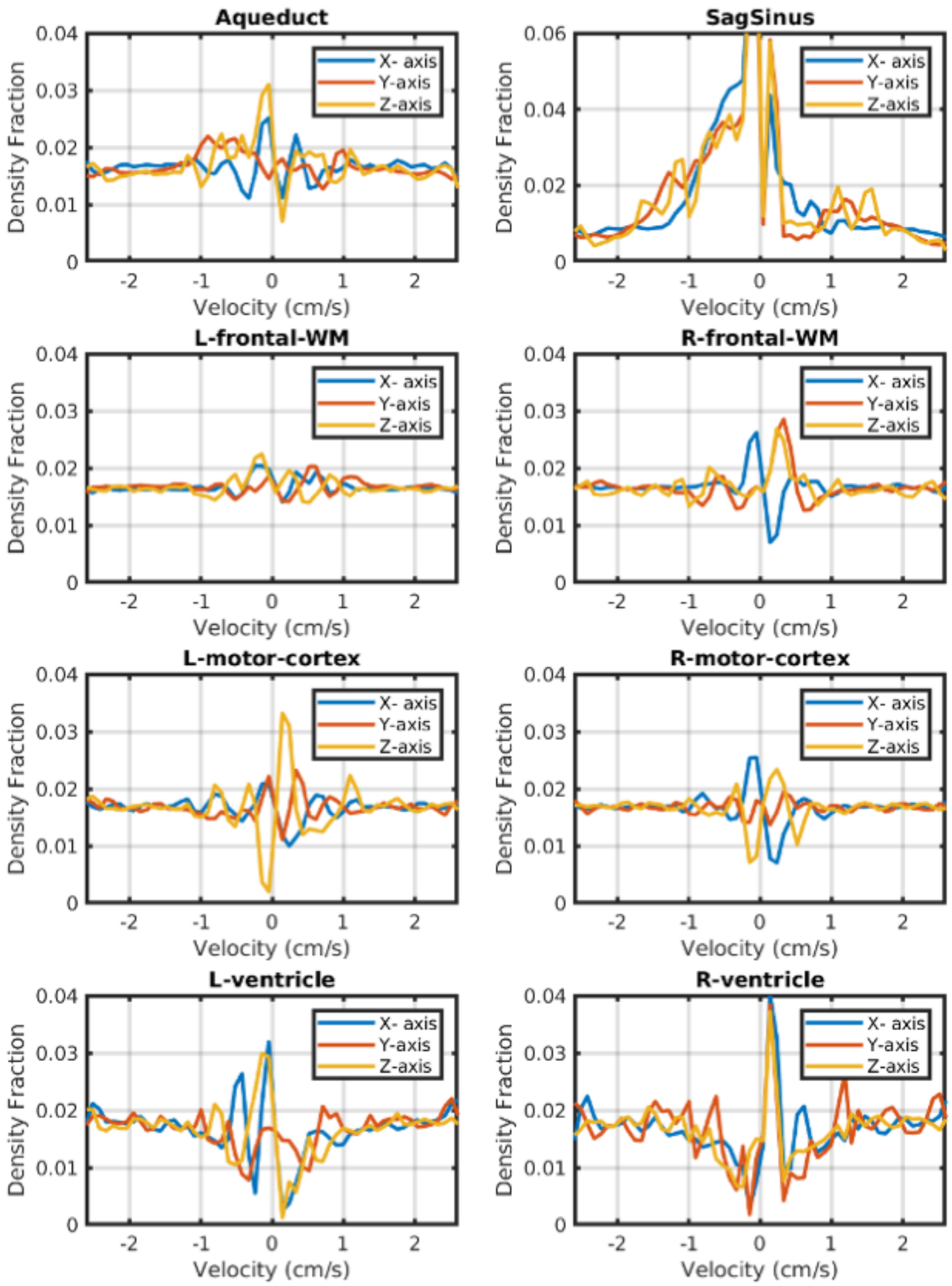

Figure 7. Velocity Spectra from 7 selected voxels extracted from one participant. The regions include the right and left motor cortices, frontal white matter regions and both frontal ventricular regions, in addition to a voxel in the cerebral aqueduct. We calculated the spatial mean was calculated for each image of the velocity spectrum image series and used it to construct a global mean spectrum regressor. We then fitted and removed this global mean series from the image series using linear regression to remove the contribution of stationary spins to the spectrum.

# Supplementary Figures

Supplemental Figure Captions (A)  Orthogonal sections of the water fraction at three different velocities (columns) along the main Cartesian axes (rows) for participant 1. The color scale indicates the fraction of spins moving at a specific velocity for each voxel. As indicated in the main text, the spatial global mean at each velocity was computed and used to regress out drift effects from the spectrum at each voxel.

(B) Single slice views of same velocity spectrum along each axis (rows). Again, the color scale indicates the fraction of spins moving at a specific velocity for each voxel.  For clarity, we only display every 5$^{th}$ (of 61) velocity bin between +/-2.175 cm/s, including the 0 cm/s velocity bin.

## Participant 1

### (A) Water Fraction at 3 velocities: Orthogonal Views

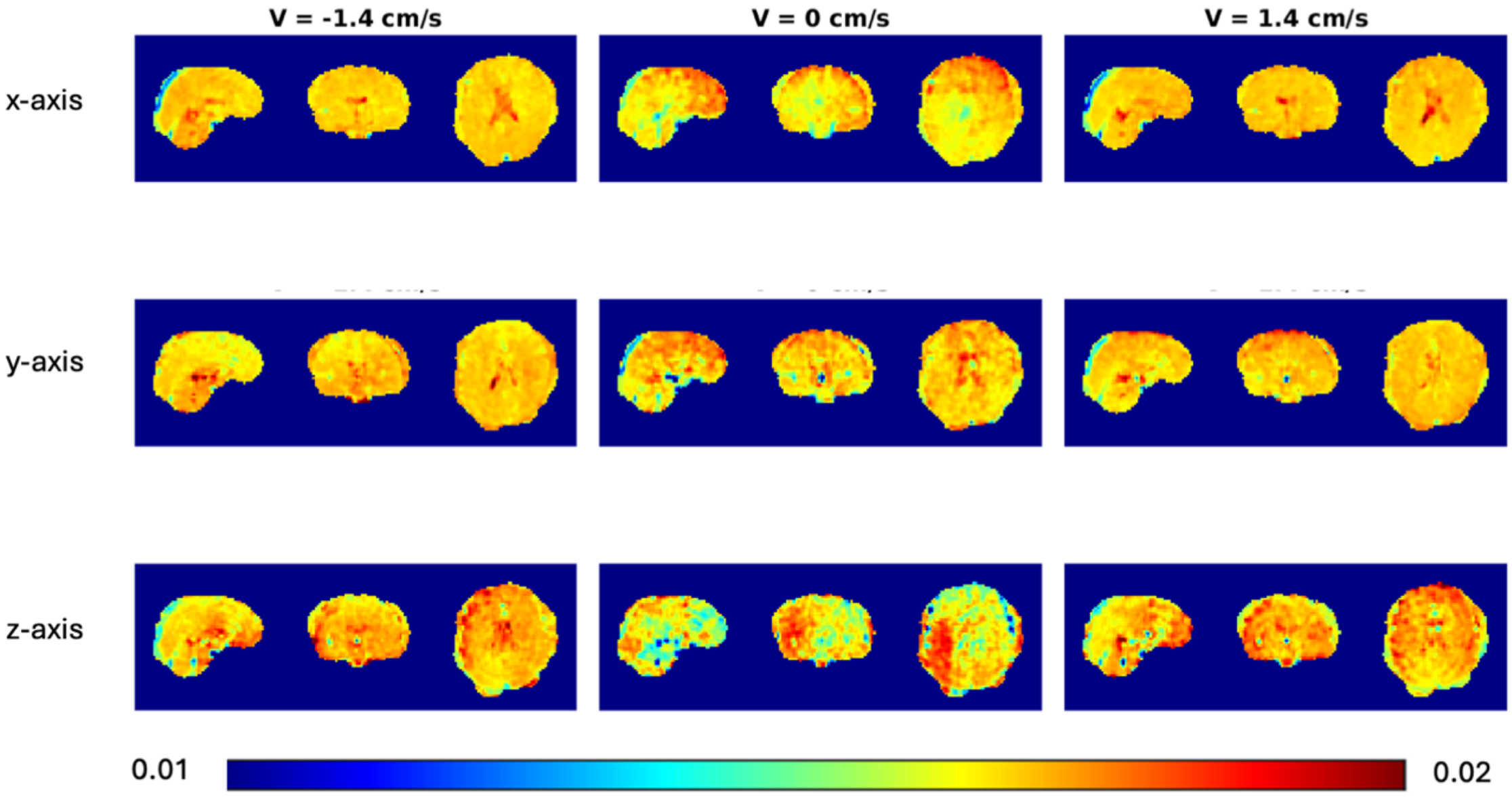

### (B) Water Fraction over Velocity Spectrum: Single Slice

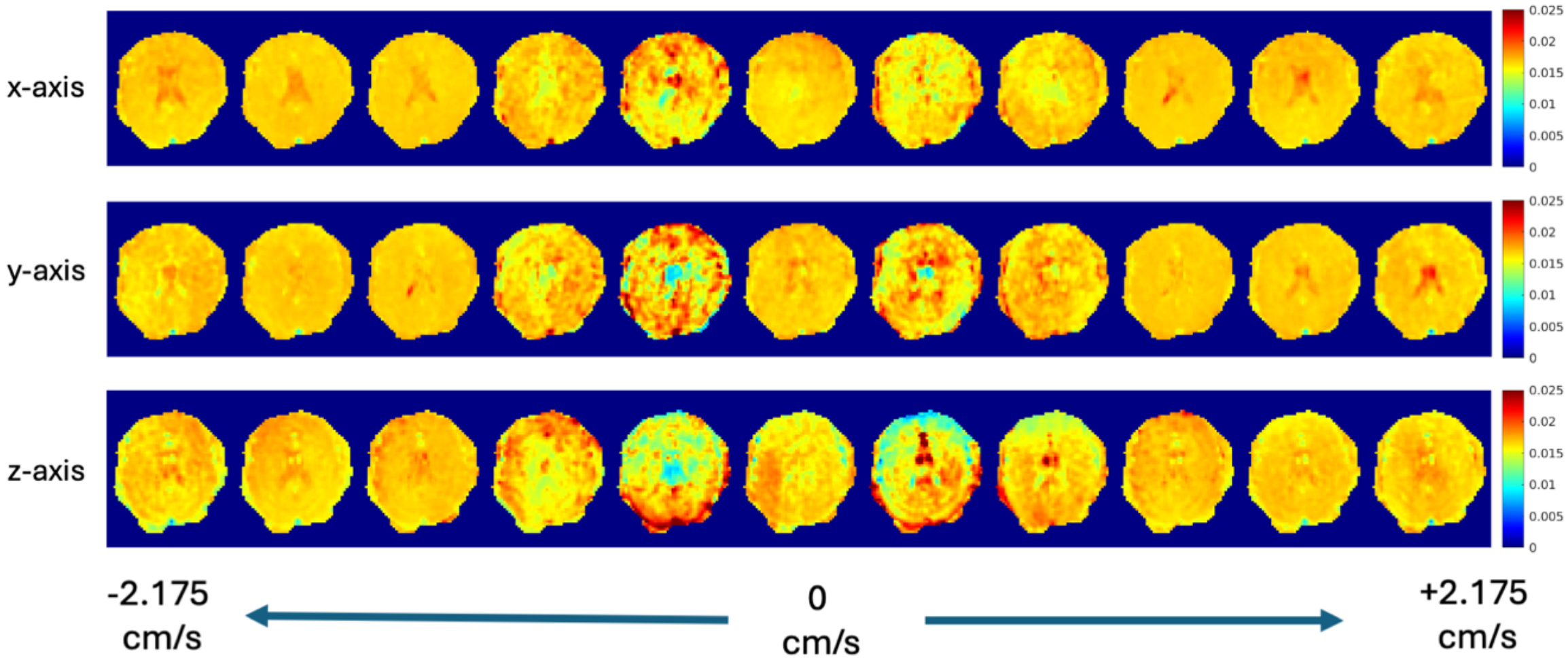

## Participant 2

### (A) Water Fraction at 3 velocities: Orthogonal Views

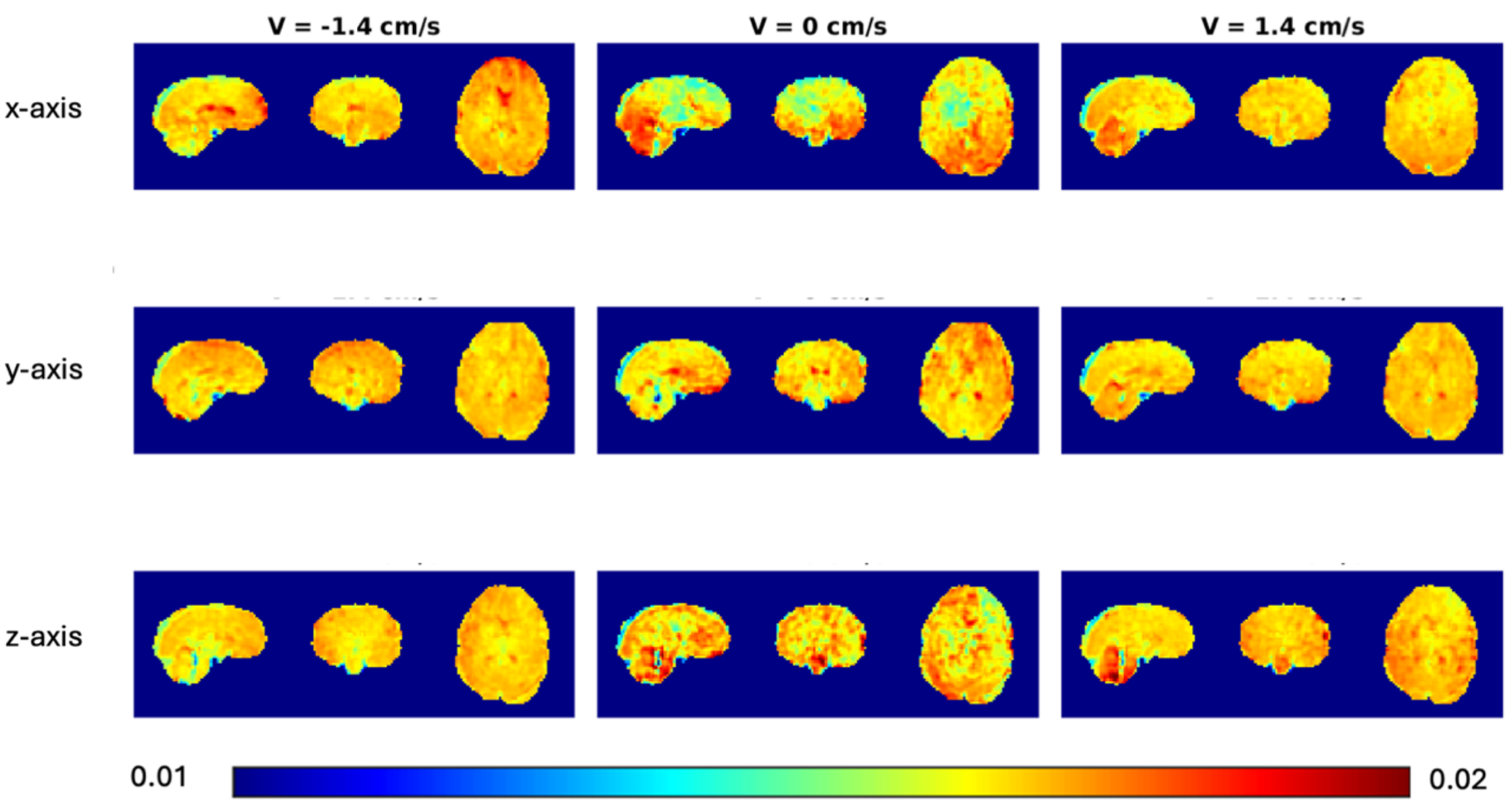

### (B) Water Fraction over Velocity Spectrum: Single Slice

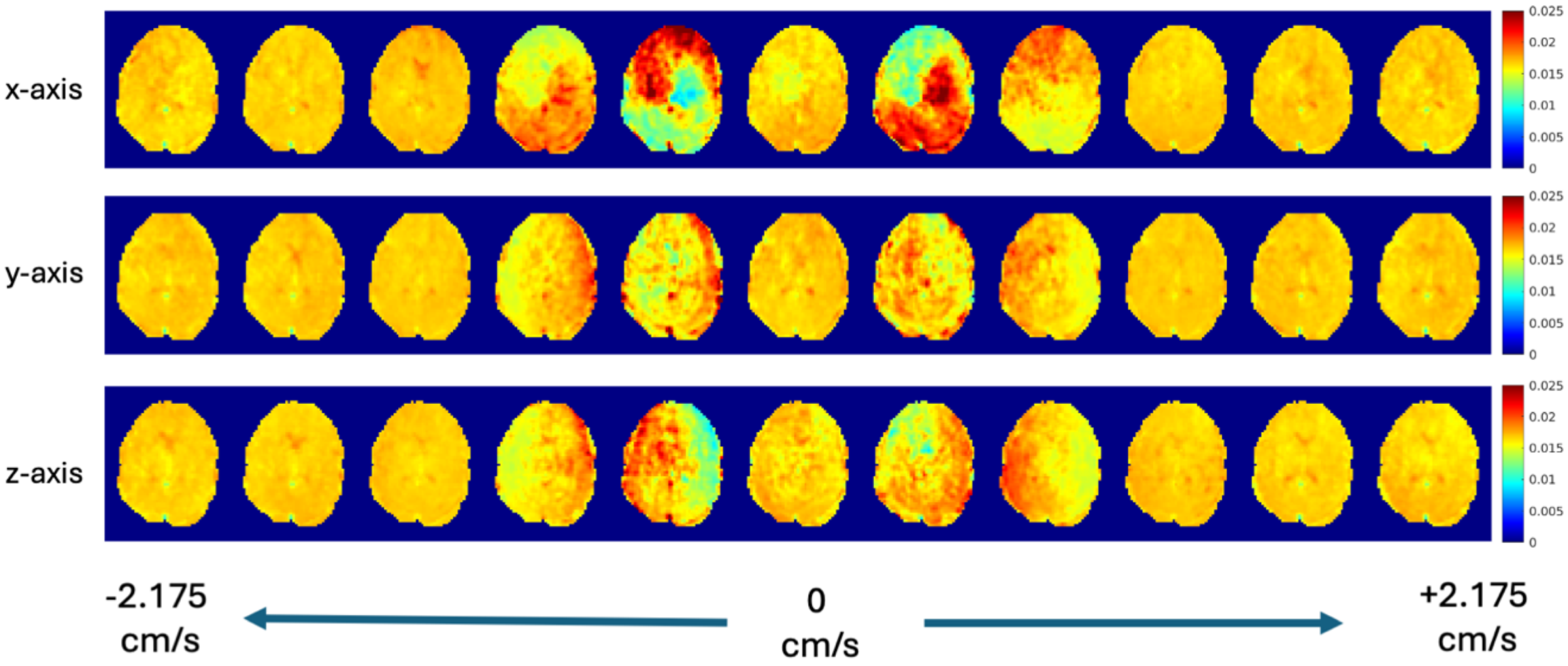

## Participant 3

### (A) Water Fraction at 3 velocities: Orthogonal Views

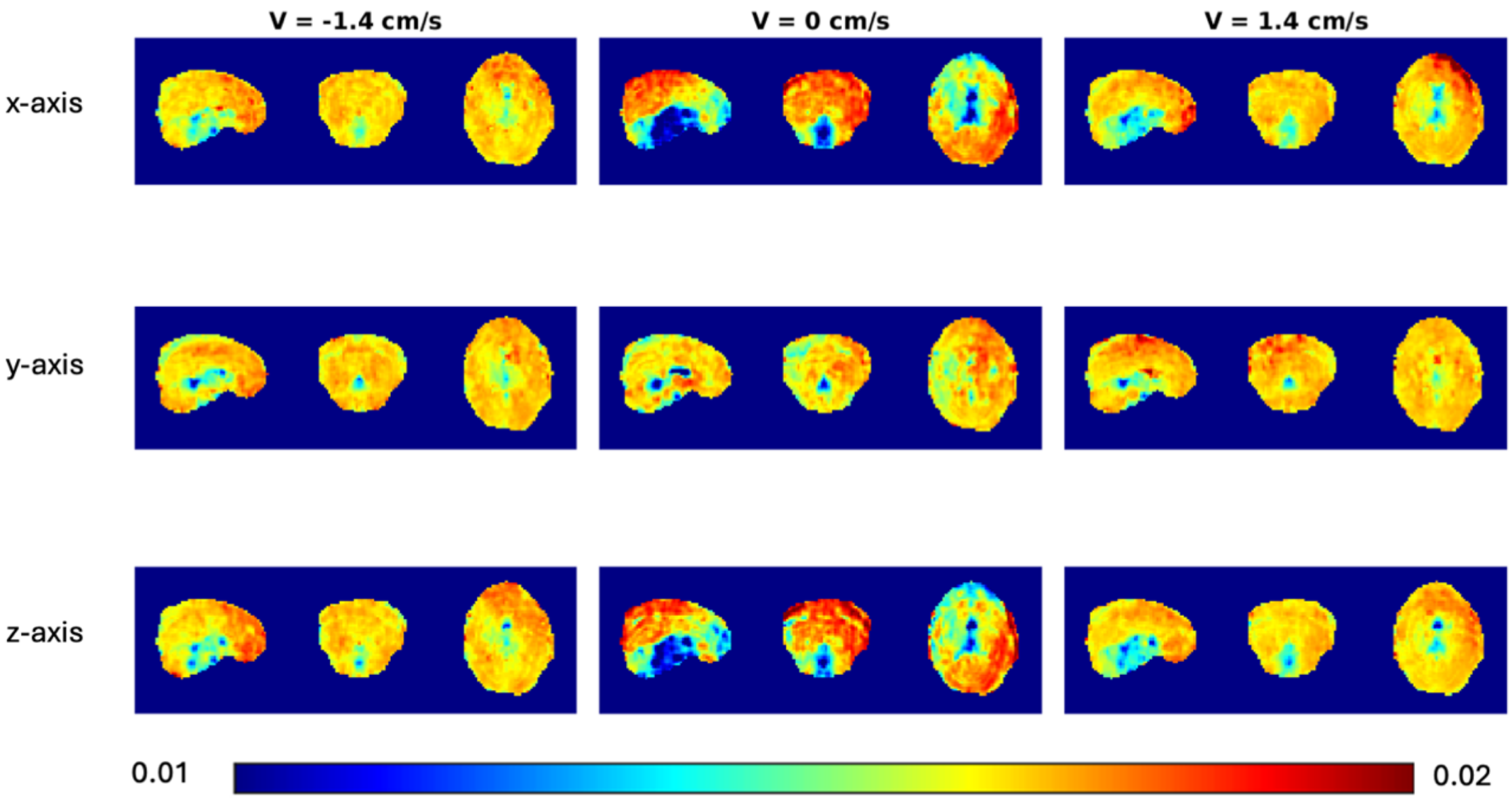

### (B) Water Fraction over Velocity Spectrum: Single Slice

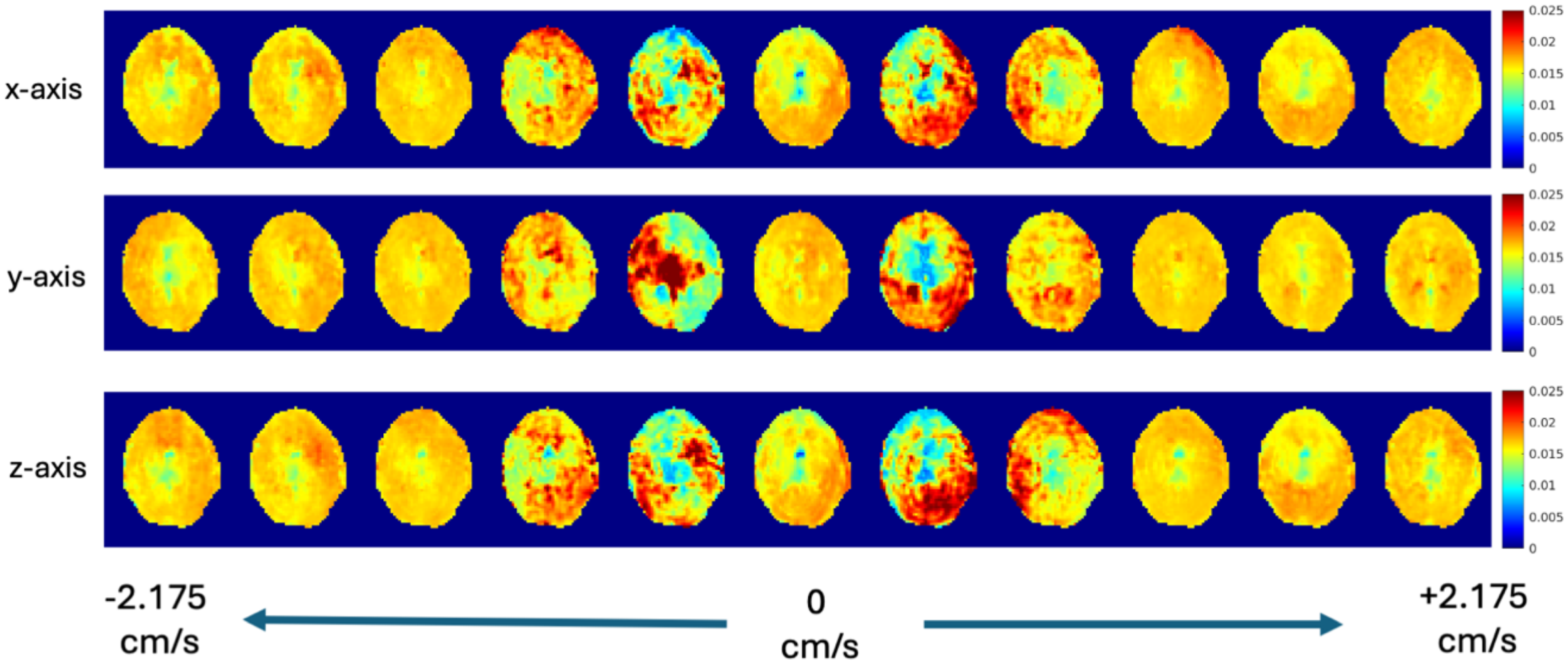

## Participant 4

### (A) Water Fraction at 3 velocities: Orthogonal Views

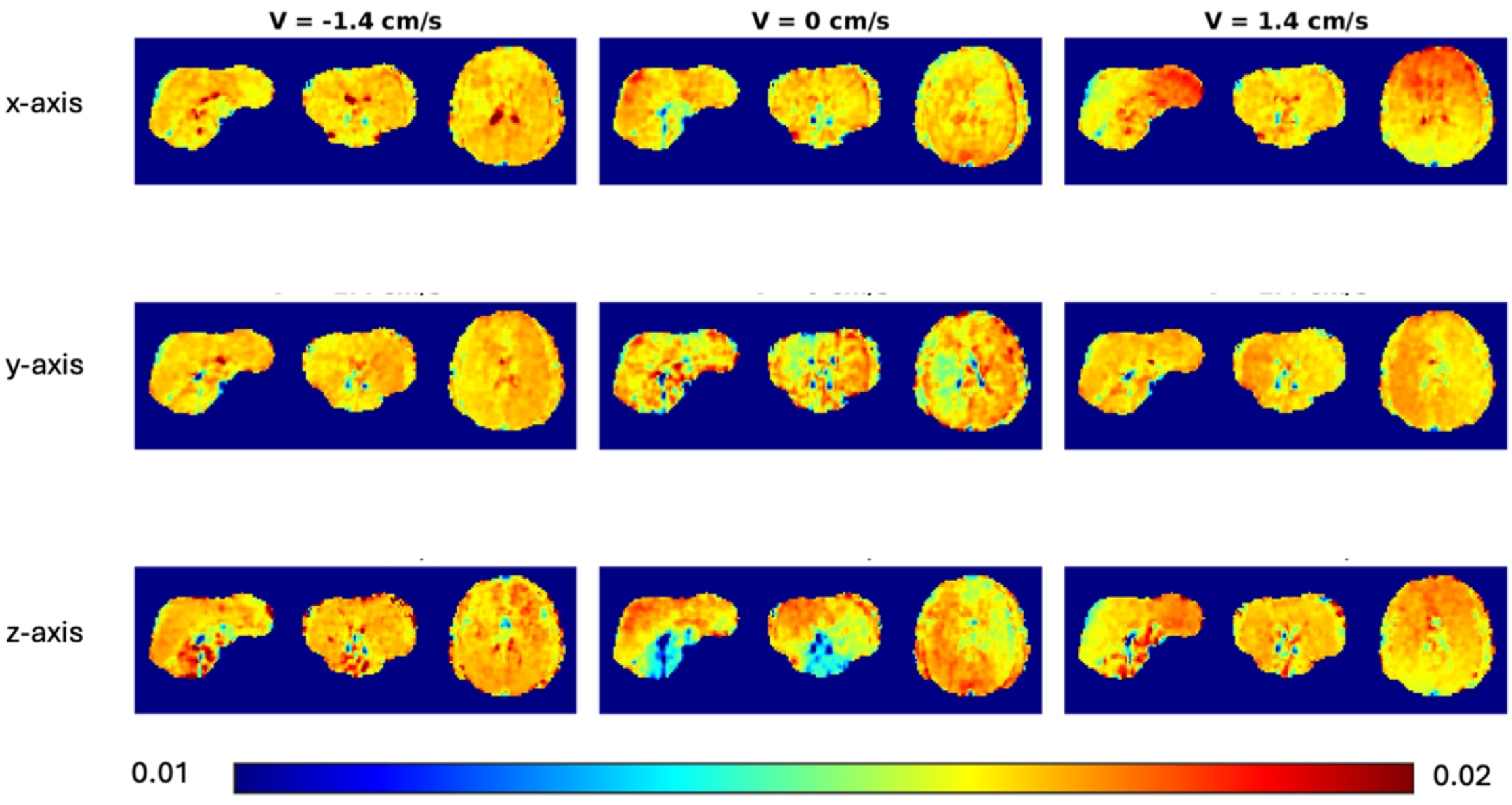

### (B) Water Fraction over Velocity Spectrum: Single Slice

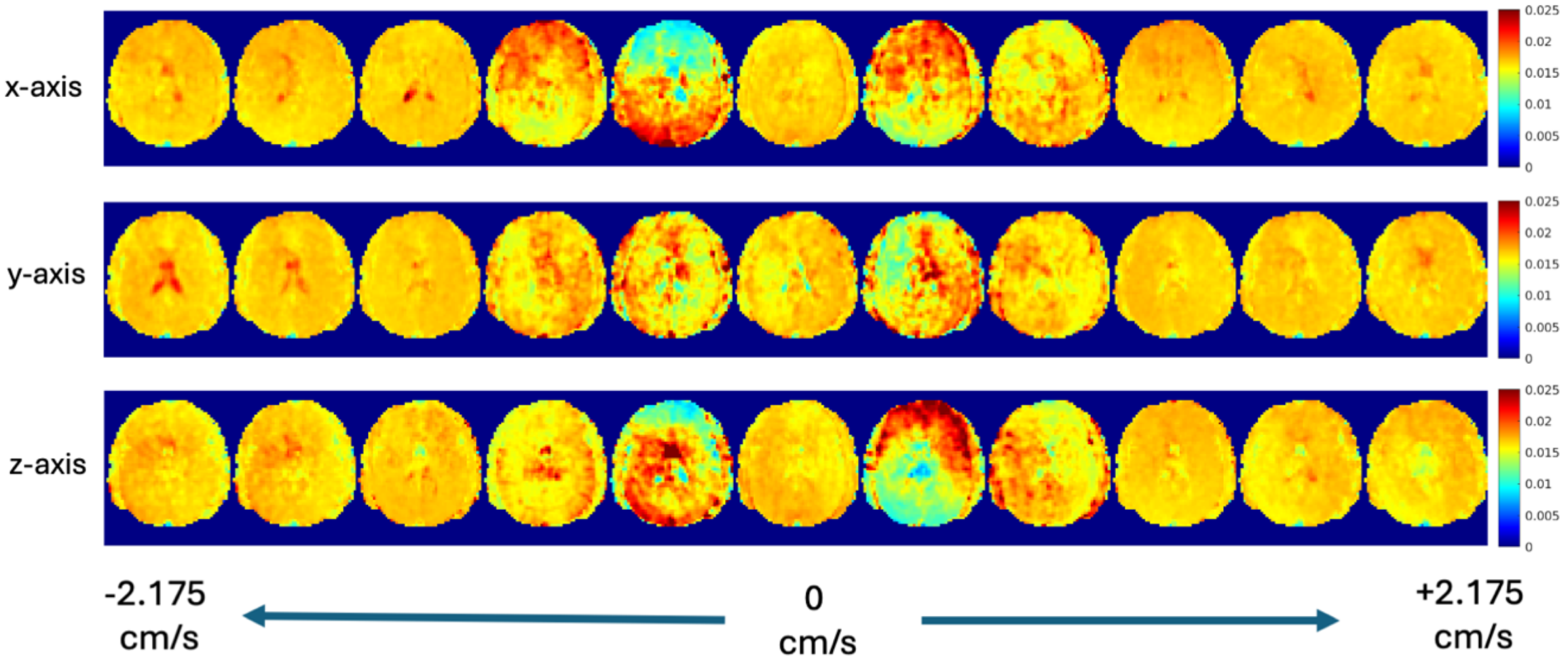

## Participant 5

### (A) Water Fraction at 3 velocities: Orthogonal Views

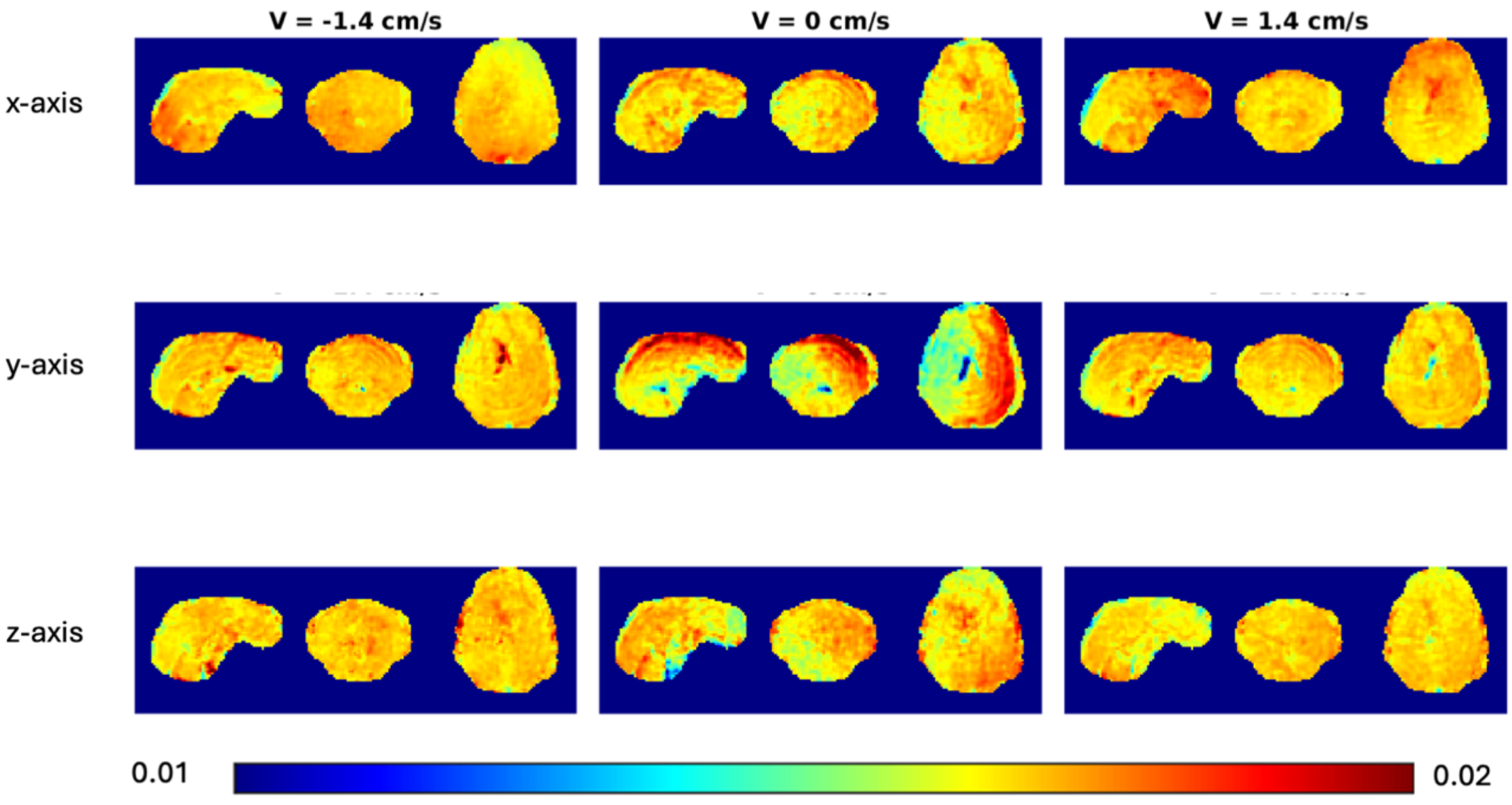

### (B) Water Fraction over Velocity Spectrum: Single Slice

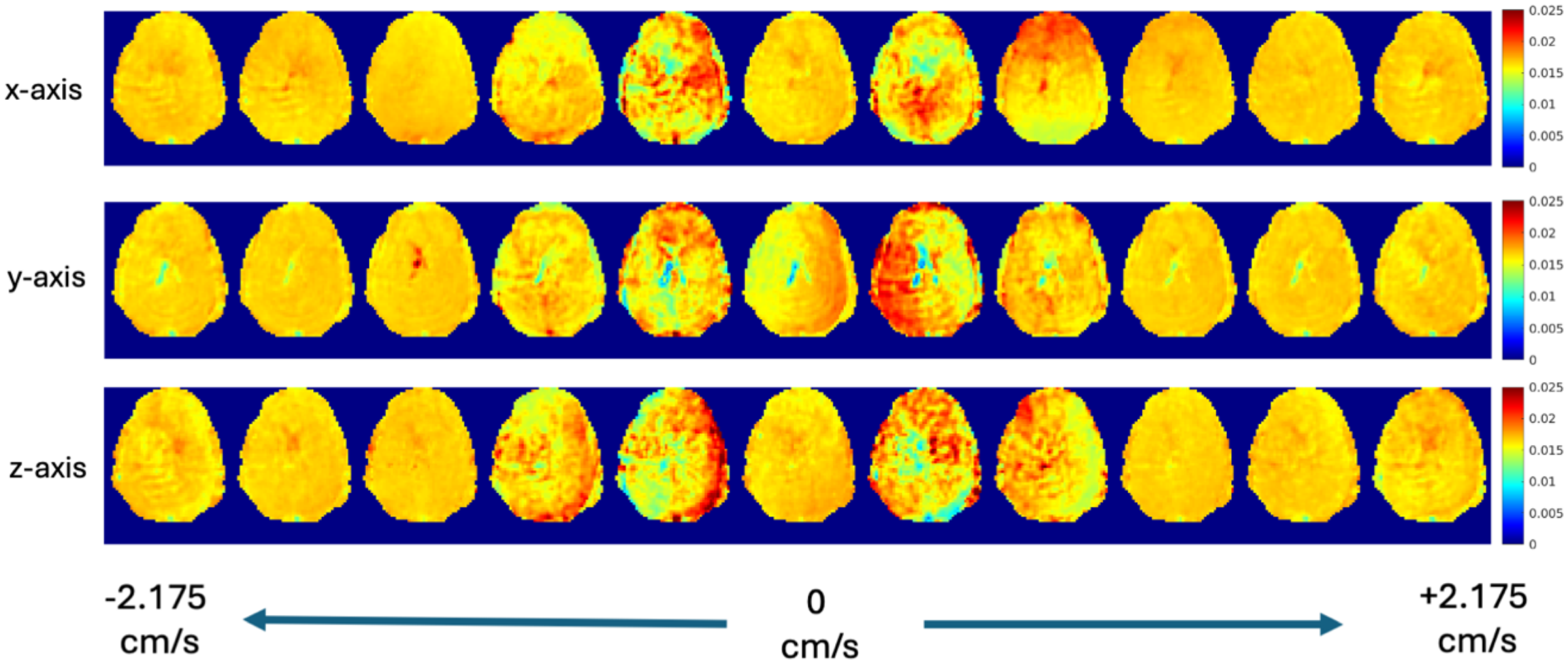